\RequirePackage{fix-cm}
\documentclass{svjour3}                     
\smartqed  
\usepackage{amsmath}
\usepackage{amsfonts}
\usepackage{amssymb}
\usepackage{setspace}
\usepackage{graphicx}
\usepackage{epstopdf}
\usepackage{cite}
\usepackage{xcolor}
\usepackage{url}
\newtheorem{algorithm}[theorem]{Algorithm}

\journalname{Sensing and Imaging}
\begin{document}

\title{Techniques in Iterative Proton CT Image Reconstruction
}

\titlerunning{Techniques in Iterative pCT Image Reconstruction}        

\author{Scott Penfold         \and
        Yair Censor }

\authorrunning{S. Penfold and Y. Censor} 

\institute{S. Penfold \at
              Department of Medical Physics\\Royal Adelaide Hospital\\Adelaide, SA 5000, Australia \\ \at
	Department of Physics\\University of Adelaide\\Adelaide, SA 5005, Australia\\
              \email{scott.penfold@health.sa.gov.au}           
           \and
           Y. Censor \at
              Department of Mathematics\\University of Haifa\\Mt. Carmel, Haifa 3498838, Israel\\
	\email{yair@math.haifa.ac.i}l
}

\date{Submitted: June 10, 2015.
	Revised: August 30th 2015.
	Revised: October 13th, 2015.}

\maketitle

\begin{abstract}
This is a review paper on some of the physics, modeling, and iterative algorithms in proton
computed tomography (pCT) image reconstruction. The primary challenge in pCT image reconstruction lies in the degraded spatial resolution resulting from multiple Coulomb scattering within the imaged object. Analytical models such as the most likely path (MLP) have been proposed to predict the scattered trajectory from measurements of individual proton location and direction before and after the object. Iterative algorithms provide a flexible tool with which to incorporate these models into image reconstruction.

The modeling leads to a large and sparse linear system of equations that can
efficiently be solved by projection methods-based iterative algorithms. Such
algorithms perform projections of the iterates onto the hyperlanes
that are represented by the linear equations of the system. They perform these
projections in possibly various algorithmic structures, such as
block-iterative projections (BIP), string-averaging projections (SAP). These
algorithmic schemes allow flexibility of choosing blocks, strings, and other
parameters. They also cater for parallel implementations which are apt to
further save clock time in computations. Experimental results are presented
which compare some of those algorithmic options.

\end{abstract}

\section{Introduction}
\label{intro}
This is a review paper on some of the physics, modeling, and iterative algorithms in proton computed tomography (pCT) image reconstruction. Iterative algorithms exist in many forms in tomographical medical imaging. The use of iterative algorithms to reconstruct images from a sequence of radiological projections dates back to the early 1970s \cite{rg70}, concurrent with the development of the first X-ray computed tomography (CT) scanner by Hounsfield \cite{houns73}. Since their initial use in X-ray CT and electron microscopy, iterative algorithms have been developed to reconstruct tomographical images in a range of modalities including single photon emission CT (SPECT), positron emission tomography (PET), magnetic resonance imaging (MRI), photoacoustic tomography, electrical impedance tomography, and recently proton CT (pCT).

Despite the long history, iterative algorithms are yet to find widespread commercial use outside the emission tomography modalities of SPECT and PET (see, e.g., \cite{xp09,eys12}). To understand why this is so, one must consider the problem in terms of physics, mathematics and computer science. Iterative reconstruction algorithms are all based on the principle of iteratively refining a discrete estimate of the reconstructed object. The refinement is based on reducing the difference between the measured projections and simulated projections based on the current image estimate. The primary distinction between the multitudes of algorithms lies in how the \textquotedblleft difference\textquotedblright $\space$ between measured and simulated projections is assessed and how the update of the current image estimate is performed. Despite the differences, there are several common challenges for iterative reconstruction algorithms:

\begin{itemize}

\item algorithms must reflect accurate modeling of the physical processes involved in the imaging modality,

\item the algorithm needs to efficiently handle large, sparse, inconsistent datasets, and

\item computation time must not be a hindrance in the clinical setting.

\end{itemize}

The first of these points justifies the full discretization approach and the use of iterative reconstruction algorithms for pCT. The advantages of such an approach over the continuous modeling and the use of analytical methods (also termed \textquotedblleft transform methods\textquotedblright) have been documented in the literature, see, e.g., \cite{CZ97}, Subsection 10.1: \textquotedblleft Transform Methods and the Fully Discretized Model" for a discussion of this topic within x-ray CT. As another example, physical modeling has played a significant role in improving the accuracy of emission tomography in nuclear medicine, particularly in the form of non-uniform attenuation correction \cite{bmt89}, see also, e.g., \cite{QL06}. In pCT, modeling of multiple Coulomb scattering (MCS) as the particle traverses the imaged object has been shown to improve spatial resolution of reconstructed images in comparison to reconstruction approaches that model proton paths as straight lines \cite{tl06}, such as the filtered backprojection (FBP) which is an analytic reconstruction method. The complexity of the physical modeling can, however, play a significant role in the speed of the reconstruction.

Memory consumption and computation time are usually seen as the challenges faced by iterative algorithms when being compared to analytical algorithms. The desire for smaller voxel sizes has come at the expense of increased memory requirements and reconstruction time. Because single projections do not involve all voxels within the reconstructed object, the datasets that are processed by the algorithm are sparse in nature. This means that the algorithm must implement efficient memory usage. Furthermore, iterative reconstructions require repeated processing of the algorithm, which can result in relatively long reconstruction times. However, with the rapidly expanding development of parallel computing environments (e.g., graphical processing units, cluster computers), clinically viable reconstruction times are a reality for physically complex imaging modalities being developed, such as pCT. Indeed, the CARP algorithm of \cite{carp}, described in Subsection \ref{subsec:carp} below, is indeed a sophisticated iterative projection method that was implemented in \cite{Karonis2013} on a cluster of CPUs and GPUs and achieved \textquotedblleft clinical\textquotedblright reconstruction times on large pCT problems.

The first tomographical reconstructions with heavy charged particle radiation were carried out with a least squares minimizing form of iterative reconstruction by Goitein in 1972 \cite{mg72}. Goitein used 2D radiographic 840 MeV alpha particle projection data measured with a body phantom to demonstrate the applicability of his algorithm. Several generations of pCT scanners were also developed between the mid 1970s and late 1990s \cite{amk76,kmh81,pz00}, with the goal of the last system being reconstruction of proton stopping powers for use in proton radiotherapy treatment planning. Although the earlier systems demonstrated desirable contrast resolution characteristics, the limited spatial resolution relative to X-ray CT and the problems associated with cost and accelerator size resulted in greater efforts being devoted to X-ray CT development. More recently however,  \cite{tl06} proposed the use of the iterative algebraic reconstruction technique (ART) \cite{rg70} to reconstruct pCT images, with the goal of improving spatial resolution by modeling the MCS process. This potential improvement in spatial resolution and the rapidly expanding use of proton and ion therapy in cancer treatment has led to renewed interest in heavy charged particle tomography. For further information regarding the development of proton radiography and pCT, the reader is referred to the recent review by Poludniowski \textit{et al.} \cite{BJR15}.

The improvement in spatial resolution when accounting for the nonlinear proton trajectory demonstrated by Li \textit{et al}. \cite{tl06} has led to the publication of several approaches to handle the proton most likely path (MLP) in pCT image reconstruction. Rit \textit{et al}. \cite{sr13} describe a distance driven FBP approach to incorporate the proton MLP. Here, binning of proton trajectories is performed at multiple depths along the MLP. A voxel-specific backprojection is then performed using the binned data from the appropriate depth in each projection angle. This approach has been shown to improve spatial resolution of pCT relative to the conventional FBP method. An alternative approach was recently suggested by Poludniowski \textit{et al}. \cite{GP14} making use of the backprojection-then-filtering (BPF) algorithm. The authors suggest this is more naturally suited to the pCT image reconstruction than FBP because of the ability to handle list-mode data and nonlinear paths inherently. While the initial results with BPF appear to be a promising direction in pCT image reconstruction, applicability to dispersive cone beam geometries has yet to be demonstrated. The above mentioned studies are based on member methods from the class of analytical image reconstruction techniques. These methods are based on creating a continuous mathematical model of the problem, solving the associated inverse problem with integral transforms and discretizing the final inversion formula. This is in contrast with iterative image reconstruction algorithms that use the fully-discretized modeling approach wherein both the object and the external source distribution are discretized at the beginning - leading to finite-dimensional vector space formulations, see, e.g., \cite{SY98}.

This article describes the implementation of iterative reconstruction algorithms for pCT, with a particular focus on projection methods for iterative reconstruction. This includes an introduction to the image reconstruction task for iterative algorithms, description of the algorithms that have been investigated for use with pCT and the incorporation of MCS into the iterative reconstruction framework.

\section{Proton CT Reconstruction in the Iterative Framework}

The goal of pCT is to reconstruct the relative stopping power (RSP) of the imaged object. The reconstructed images can then be used in treatment planning to more accurately calculate dose deposition with treatment beams. To reconstruct RSPs, measurements of proton energy loss upon traversal of the imaged object are required. If the energy loss is recorded for $I$ individual protons and we wish to reconstruct the RSP in a square matrix containing $J$ pixels, the discretized problem statement becomes

\begin{equation}
A x = b,
\label{eq:prob}
\end{equation}

\noindent where $A$ is the $I \times J$ \textit{system matrix} containing elements $a^i_j$ describing the length of intersection of the $i$-th proton with the $j$-th pixel, $x$ is a $J$-dimensional vector containing the estimated RSPs, and $b$ is an $I$-dimensional vector containing the water equivalent pathlengths (WEPLs) of the $I$ proton measurements. Assuming the elements of $A$ can be determined from the proton tracking measurements and the MLP formalism \cite{dcw04,rws08} and $b$ from the energy loss measurements, the image reconstruction task in pCT is to determine the RSP vector $x$.

Some practical considerations with regard to the calculation of $A$ and $b$ should be noted. To accurately calculate the elements of the $A$ matrix one must know where in the reconstruction region multiple Coulomb scattering took place. This requires an estimate of the patient contour. Several approaches to this problem have been presented including the use of the FBP as an initial estimate and hull-detection algorithms \cite{Schultze14}. It should also be noted that although the problem statement was made with respect to a 2D geometry, it is equally applicable to 3-dimensional reconstructions where the 3D scattered proton path is projected onto two perpendicular 2D planes and the length of intersection of the proton paths with 3D voxels is calculated. The 2D projections only become invalid if a proton travels parallel to the third dimension. However, since this situation will require large angle nuclear scattering interactions to take place, it is likely that the large angle scattering and/or large energy loss processing will remove these proton histories from the reconstruction. 

The elements of the system matrix also depend on the choice of basis functions for the fully discretized model. In fact, they need not be limited to a square pixel representation, and the use of radially symmetric basis functions may result in more favourable reconstructions (see, \cite{rml92,kmh85}). Finally, to obtain the elements of the $b$ vector, the WEPLs can either be analytically derived from the proton energy loss measurements, or directly obtained from an appropriately calibrated detector \cite{Hurley12}.

\section{Iterative Reconstruction Families\label{sec:iterative}}

\subsection{Statistical Iterative Reconstruction \label{sec:statistical}}

Data collected for medical imaging is inevitably subject to noise and can therefore cause the problem statement in Equation (\ref{eq:prob}) to be inconsistent. Because radiation interactions are stochastic in nature, inconsistent data are unavoidable. This physical process is a fundamental component of \textquotedblleft salt and pepper\textquotedblright\ noise in the reconstructed image. Statistical iterative reconstruction algorithms were applied to medical imaging with the goal of reducing noise in the reconstructed image by physically modeling the processes giving rise to inconsistent data.

The first application of Poisson likelihood models to both emission and transmission forms of image reconstruction were provided by Rockmore and Macovski \cite{ajr76,Rock77}. The original goal of these applications was to reduce the statistical noise present in gamma emission tomography when images were reconstructed with an analytical filtered backprojection approach \cite{ajr76}. This application was successful because of the relatively large photon counting distribution (Poisson in nature) experienced in emission tomography modalities such as PET and SPECT. Although superior reconstruction results were achieved, the long computation times meant that statistical iterative reconstruction was not immediately applied to the clinical setting. However, with advances in computing technology, the class of expectation maximization (EM) statistical iterative reconstruction algorithms \cite{Shepp82} eventually found widespread use in clinical emission tomography.

Despite the success in PET and SPECT, statistical iterative reconstruction algorithms are relatively uncommon in transmission tomography (X-ray CT or pCT). The primary reason for this is the much greater signal-to-noise ratio achieved in transmission tomography. Therefore, X-ray CT images reconstructed with analytical algorithms do not suffer to the same extent from statistical noise as those in PET and SPECT. However, with growing interest in low-dose imaging, radiation intensities used in transmission scanning may be significantly reduced, leading to a need for statistical algorithms to model this noisy data.

While photon counting (which is the basis for transmission and emission tomography) may be well modelled by Poisson statistics, modern pCT is based on single particle energy measurements. Variation in energy measurements is termed energy straggling and is primarily a result of the stochastic nature of proton energy loss in a single interaction process.

When a heavy charged particle traverses a medium that is thin compared to the range of the particle in that medium, the energy loss distribution resulting from stochastic energy loss can be described by a Vavilov distribution \cite{pvv57}. Fortunately, for thick objects, which a human patient can be considered to be, the energy loss distribution can be modelled by the well-known Gaussian function. This observation suggests that statistical iterative reconstruction may be used to reduce the statistical noise in pCT images resulting from energy loss straggling. To date there have been no applications of statistical iterative reconstructions in the field of heavy ion or pCT with the purpose of modeling straggling induced noise. In the following text a simplified example is given to demonstrate how such an algorithm might be applied to account for the inherent distribution of energy loss values.

We begin with the assumption that for thick absorbing materials, the energy loss distribution of energetic charged particles is well described by a Gaussian distribution. While a large energy loss tail also exists, this component can be rejected from a pCT reconstruction by analysis of the energy loss of proton histories that shared a similar spatial trajectory. If the measured WEPL is assumed to be the mean about which the calculated WEPL is distributed and standard deviation of the Gaussian distribution $\sigma$ is known, the likelihood $L$ that the estimated stopping power resulted in the measured WEPL for the $i$-th proton is given by%
\begin{equation}
L \left( \sum_j a^i_j x_j \bigg| b_i, \sigma \right) = \frac{1}{\sqrt{2 \pi} \sigma} \exp \left( - \frac{ \left( \sum_j a^i_j x_j - b_i \right) ^2}{2 \sigma^2} \right).
\end{equation}

In statistical iterative analysis, the log-likelihood is often used in place of the likelihood for mathematical convenience. In the case of the Gaussian function, the log-likelihood is simply

\begin{equation}
\ell \left( \sum_j a^i_j x_j \bigg | b_i \right) = \left( \sum_j a^i_j x_j - b_i \right) ^2
\label{eq:loglikelihood}
\end{equation}

\noindent when all constants are ignored. This is justified because the constants will have the same effect on the log likelihood for all estimates of the image vector $x$. In the statistical iterative reconstruction framework, this log-likelihood is known as the cost function. It is clear from Equation (\ref{eq:loglikelihood}) that the objective of the reconstruction is to minimize this cost function. When idealized noise-free data is used, the cost function will reduce to zero.

In the case of Gaussian distributed data, the image reconstruction problem reduces to an instance of the well-known method of least squares. A form of this method was applied by Goitein in the first reconstructions of heavy-charged particle transmission data in 1972 \cite{mg72}. Since then, only one recent publication has investigated the application of statistical algorithms to pCT image reconstruction \cite{Lee2015}. The authors compared expectation maximization (EM) and projection based iterative algorithms for a prototype pCT scanner. The projection based algorithms showed superior quantitative reconstruction results, however, the EM algorithm showed superior convergence rates. The projection based methods have been more widely investigated and are described in greater detail in the following sections.

Despite the lack of studies, further investigation into statistical iterative reconstruction may prove to be an effective method for reducing straggling induced noise in pCT reconstruction. The reader is referred to \cite{ps00,jf02} for further details on statistical iterative reconstruction algorithms such as introducing regularization parameters and implementing the iterative algorithms used to minimize the cost function.

\subsection{Projection Methods for Iterative Reconstruction\label{sec:algebraic}}

Several recently investigated iterative pCT reconstruction algorithms belong to the class of \textit{projection methods}. This family of iterative reconstruction algorithms will be introduced in this section by reviewing algorithmic structures and specific algorithms. Projection methods employ projections onto convex sets. They can solve a variety of feasibility-seeking or optimization problems. With different algorithmic structures, of which some are particularly suitable for parallel computing, they demonstrate nice convergence properties and/or good initial numerical behavior patterns. This class of algorithms has witnessed great progress in recent years and its member algorithms have been applied with success to fully-discretized models in image reconstruction and image processing, see, e.g., \cite{SY98}, \cite{CZ97}, and the recent \cite{cccdh10}.

The \textit{convex feasibility problem }is to find a point (any point) in the non-empty intersection $C:=\cap_{i=1}^{I}C_{i}\neq\emptyset$ of a family of closed convex subsets $C_{i}\subseteq R^{J}$, $1\leq i\leq I,$ of the $J$-dimensional Euclidean space. It is a fundamental problem in many areas of mathematics and the physical sciences, see, e.g., \cite{c93,c96}. It has been used to model significant real-world problems in image reconstruction from projections, see, e.g., \cite{H80}, in radiation therapy treatment planning, see, e.g., \cite{censor03}, and has been used in other fields under additional names such as \textit{set theoretic estimation} or the \textit{feasible set approach}. A common approach to such problems is to use projection algorithms, see, e.g., \cite{bb96} and the recent \cite{bk13}, which employ \textit{orthogonal projections} (i.e., nearest point mappings) onto the individual sets $C_{i}.$ The orthogonal projection $P_{\Omega}(z)$ of a point $z\in R^{J}$ onto a closed convex set $\Omega\subseteq R^{J}$ is defined by%
\begin{equation}
P_{\Omega}(z):=\operatorname{argmin}\{\parallel z-x\parallel_{2}\mid\text{ }x\in\Omega\},
\end{equation}
where $\parallel\cdot\parallel_{2}$ is the Euclidean norm in $R^{J}.$ Frequently, a \textit{relaxation parameter} $\lambda$ is introduced so that%
\begin{equation}
P_{\Omega,\lambda}(z):=(1-\lambda)z+\lambda P_{\Omega}(z)\label{eq2}%
\end{equation}
is the \textit{relaxed projection} of $z$ onto $\Omega$ with relaxation $\lambda.$ Since linear equations, represented by hyperplanes, or linear inequalities, represented by half-spaces, are common in pCT reconstruction we give the following expressions for the projections onto them. Let%
\begin{equation}
H=\{x\in R^{J}\mid\left\langle a,x\right\rangle =b\}
\end{equation}
be a hyperplane that represents the linear equation $\left\langle a,x\right\rangle =b$ where $a=(a_{j})_{j=1}^{J}\in R^{J}$ is a given vector, $\left\langle a,x\right\rangle =\sum_{j=1}^{J}a_{j}x_{j}$ is the inner product of $a$ and $x=(x_{j})_{j=1}^{J}\in R^{J}$ and $b$ is a given real number. Then the projection of a point $z=(z_{j})_{j=1}^{J}\in R^{J}$ onto $H$ is%
\begin{equation}
P_{H}(z)=z+\frac{b-\langle a,z\rangle}{\Vert a\Vert_{2}^{2}}a.
\end{equation}
For a half-space%
\begin{equation}
G=\{x\in R^{J}\mid\left\langle a,x\right\rangle \leq b\}
\end{equation}
that represents the linear inequality $\left\langle a,x\right\rangle \leq b,$ the projection of a point $z=(z_{j})_{j=1}^{J}\in R^{J}$ onto $G$ is%
\begin{equation}
P_{G}(z)=\left\{
\begin{array}
[c]{ll}%
z+\frac{\displaystyle b-\displaystyle\langle a,z\rangle}{\displaystyle\Vert
a\Vert_{2}^{2}}a, & \text{if }\left\langle a,z\right\rangle >b,\\
z, & \text{if }\left\langle a,z\right\rangle \leq b.
\end{array}
\right.
\end{equation}
\qquad\

\section{Sequential and Simultaneous Iterative Algebraic Reconstruction\label{sec:seq-and-sim-algs}}

The well-known Algebraic Reconstruction Technique (ART), see \cite{H80} for details and references, is a \textit{sequential} projection algorithm. Starting from an arbitrary initial point $x^{0}\in R^{J},$ the algorithm's iterative step is%
\begin{equation}
x^{k+1}=x^{k}+\lambda_{k}(P_{C_{i(k)}}(x^{k})-x^{k}),\label{alg:pocs}%
\end{equation}
where the sets $C_{i}$ can be either hyperplanes or half-spaces so that $P_{C_{i(k)}}$ is either $P_{H_{i(k)}}$ or $P_{G_{i(k)}},$ respectively. The $\{\lambda_{k}\}_{k=0}^{\infty}$ are relaxation parameters and $\{i(k)\}_{k=0}^{\infty}$ is a \textit{control sequence}, $1\leq i(k)\leq I,$ for all $k\geq0,$ which determines the individual set $C_{i(k)}$ onto which the current iterate $x^{k}$ is projected. 

A commonly used control is the \textit{cyclic control }in which $i(k)=k\operatorname*{mod}I+1,$ but other controls are also available, see, e.g., \cite[Definition 5.1.1]{CZ97}. When formulated for convex sets, as in (\ref{alg:pocs}), it has been named in the literature \textquotedblleft Projections Onto Convex Sets\textquotedblright\ (POCS) algorithm for the convex feasibility problem, see, e.g., \cite{y87}; for hyperplanes it is recognized as the Kaczmarz algorithm \cite{Kczmarz}. The origins of the POCS method are in \cite{b65,gpr67}.

In contrast with the way ART uses the constraints sequentially, one at a time in each iteration, the simultaneous counterpart of ART uses all equations (or inequalities, or convex constraints sets) in each iterative step. Starting from an arbitrary initial point $x^{0}\in R^{J},$ the algorithm's iterative step is%
\begin{equation}
x^{k+1}=x^{k}+\lambda_{k}\left(  \sum_{i=1}^{I}w_{i}\left(  P_{C_{i}}%
(x^{k})-x^{k}\right)  \right)  ,\label{eq:simultaneous}%
\end{equation}
where for hyperplanes or half-spaces $P_{C_{i}}$ is either $P_{H_{i}}$ or $P_{G_{i}},$ respectively. The $\{\lambda_{k}\}_{k=0}^{\infty}$ are relaxation parameters, and $\{w_{i}\}_{i=1}^{I}$ are positive weights such that $\sum_{i=1}^{I}w_{i}=1.$ For linear equations this algorithm has been first published by Cimmino \cite{Cimmino} and is thus sometimes called after him.

\section{Block-Iterative and String-Averaging Iterative Reconstruction\label{sec:bip-and-sap}}

The sequential and simultaneous projection algorithmic schemes described above are in fact special cases of the more general schemes of \textit{block-iterative projections }(BIP) and of \textit{string-averaging projections} (SAP) methods. A classification of projection algorithms into such classes appears in \cite[Section 1.3]{CZ97}, and the review paper \cite{bb96} presents a variety of specific algorithms of these kinds, while the more recent SAP can be found in \cite{cz12} and references therein. The structural algorithmic skeleton of BIP and SAP methods is similar to what is known in other fields as: \textquotedblleft block-sequential\textquotedblright and \textquotedblleft block-parallel\textquotedblright schemes, see, e.g., \cite{wk80}.

\subsection{The Block-Iterative Projections Algorithmic Structure\label{subsec:bip}}

The \textit{block-iterative projections} (BIP) algorithmic scheme is presented here for fixed blocks, although, as proposed in \cite{ac89}, it can accommodate variable blocks and variable weight systems. It starts with the creation of \textquotedblleft blocks\textquotedblright. For $t=1,2,\dots,T,$ the \textit{block} $B_{t}$ is a subset of $\{1,2,\dots,I\}$ of the form%
\begin{equation}
B_{t}=(i_{1}^{t},i_{2}^{t},\dots,i_{n(t)}^{t}),
\end{equation}
with $n(t)$ denoting the number of elements in $B_{t}.$ Since we restrict our description to fixed blocks, every index must be contained in one of the blocks but the blocks need not be disjoint. With each block $B_{t}$ a weights system is defined by choosing weights $\{w_{i}^{t}\}_{i\in B_{t}}$ which are positive real numbers $w_{i}^{t}>0$ such that $\sum_{i\in B_{t}}w_{i}^{t}=1.$ Initializing the algorithm at an arbitrary $x^{0}\in R^{J},$ in the iterative step of the BIP algorithmic scheme, when the current iterate $x^{k}$ is available, a control index $t(k)$ is picked, where $1\leq t(k)\leq T,$ according to some rule, say, the cyclic rule of $t(k)=k\operatorname*{mod}T+1$ (other controls are also available). Then the algorithm performs a fully-simultaneous iteration, like that of (\ref{eq:simultaneous}), with respect to only the constraints sets whose indices are in the chosen block $B_{t(k)}$ by the iterative step%
\begin{equation}
x^{k+1}=x^{k}+\lambda_{k}\left(  \sum_{i\in B_{t(k)}}w_{i}^{t(k)}\left(
P_{C_{i}}(x^{k})-x^{k}\right)  \right)  ,
\end{equation}
where for hyperplanes or half-spaces $P_{C_{i}}$ is either $P_{H_{i}}$ or $P_{G_{i}},$ respectively, and the $\{\lambda_{k}\}_{k=0}^{\infty}$ are relaxation parameters.

In this framework we obtain a sequential projection algorithm like (\ref{alg:pocs}) by the choice $T=I$ and $B_{t}=(t)$ for all $t=1,2,\dots,T,$ and a simultaneous projection algorithm like (\ref{eq:simultaneous}) by the choice $T=1$ and $B_{1}=(1,2,\dots,I)$. See also \cite{Elfving,ehl81}.

\subsection{The String-Averaging Projections Algorithmic Structure\label{subsec:sap}}

The \textit{string-averaging projections} (SAP) algorithmic scheme (first proposed in \cite{ceh01}) starts with the creation of \textquotedblleft strings\textquotedblright. For $t=1,2,\dots,T,$ the \textit{string} $I_{t}$ is an ordered subset of $\{1,2,\dots,I\}$ of the form%
\begin{equation}
I_{t}=(i_{1}^{t},i_{2}^{t},\dots,i_{m(t)}^{t}),\label{block}%
\end{equation}
with $m(t)$ denoting the number of elements in $I_{t}.$ Since we restrict our description to fixed strings, every index must be contained in one of the strings but the strings need not be disjoint. Initializing the algorithm at an arbitrary $x^{0}\in R^{J},$ the iterative step of the SAP algorithmic scheme generates from the current iterate $x^{k}$, for all $t=1,2,\dots,T,$ the following intermediate points that are obtained by sequentially projecting $x^{k}$ onto the sets whose indices belong to the $t$-th string $I_{t},$ namely,
\begin{equation}
S_{t}(x^{k})=P_{i_{m(t)}^{t}}\cdots P_{i_{2}^{t}}P_{i_{1}^{t}}(x^{k}%
),\label{eq:strings}%
\end{equation}
where $P_{i_{r}^{t}}$ denotes the projection $\displaystyle P_{C_{i_{r}^{t}}}$\ onto the set $C_{i_{r}^{t}}$ which for hyperplanes or half-spaces is either $\displaystyle P_{H_{i_{r}^{t}}}$ or $\displaystyle P_{G_{i_{r}^{t}}},$ respectively, for $1\leq r\leq m(t)$.\ This can be done in parallel for all strings. Then a convex combination of the strings' end-points $S_{t}(x^{k})$, for all $t=1,2,\dots,T,$ is calculated by%
\begin{equation}
x^{k+1}=\sum_{t=1}^{T}w_{t}S_{t}(x^{k})\label{as2}%
\end{equation}
where $\{w_{t}\}_{t=1}^{T}$ are positive weights such that $\sum_{t=1}^{T}w_{t}=1.$

In this framework we obtain a sequential projection algorithm like (\ref{alg:pocs}) by the choice $T=1$ and $I_{1}=(1,2,\dots,I)$ and a simultaneous projection algorithm like (\ref{eq:simultaneous}) by the choice $T=I$ and $I_{t}=(t),\ $for all $t=1,2,\dots,T.$

We demonstrate the underlying idea of the BIP and SAP algorithmic schemes with the aid of Figure 1. Figure 1(a) depicts the fully sequential ART (Kaczmarz) algorithm and the fully simultaneous (Cimmino) algorithm appears in Figure 1(b). In Figure 1(c) we show how a simple averaging of successive projections (as opposed to averaging of parallel projections as in Figure 1(b)) works. In this case $T=I$ and $I_{t}=(1,2,\dots,t)$, for $t=1,2,\dots,T.$ This scheme, appearing in \cite[Example 2.14]{bb96}, inspired the formulation of the general string-averaging algorithmic scheme whose action is demonstrated in Figure 1(d). It averages, via convex combinations, the end-points obtained from strings of sequential projections and in this figure the strings are $I_{1}=(1,3,5,6),$ $I_{2}=(2),$ $I_{3}=(6,4)$. 

BIP and SAP algorithmic schemes offer a variety of options for steering the iterates towards a solution of the convex feasibility problem. They are \textit{inherently parallel} schemes in that their mathematical formulation is parallel as opposed to algorithms which are sequential in their mathematical formulation but can, sometimes, be implemented in a parallel fashion based on appropriate model decomposition (i.e., depending on the structure of the underlying problem). Inherently parallel schemes enable flexibility in the actual manner of implementation on a parallel machine. This is of particular importance in pCT where calculation of individual proton MLPs is a computationally expensive process. If multiple MLPs can be computed in parallel with an algorithm executed on a multi-processor architecture, image reconstruction times may be significantly reduced \cite{Karonis2013}.
\begin{figure}[t]
\[
\begin{array}
[c]{cc}
{\includegraphics[
height=1.8568in,
width=2.0358in
]
{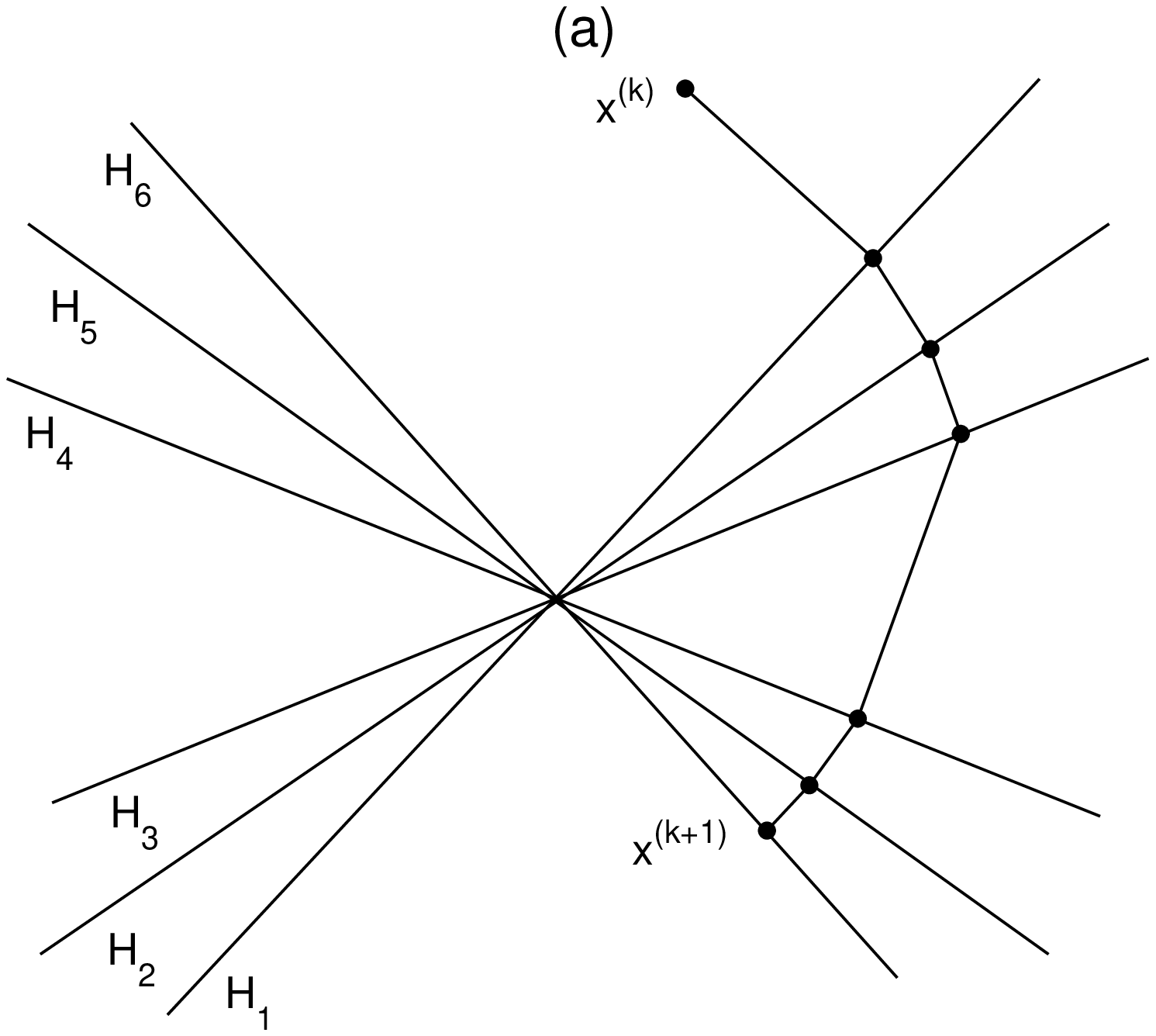}%
}
&
{\includegraphics[
height=1.8568in,
width=2.0358in
]
{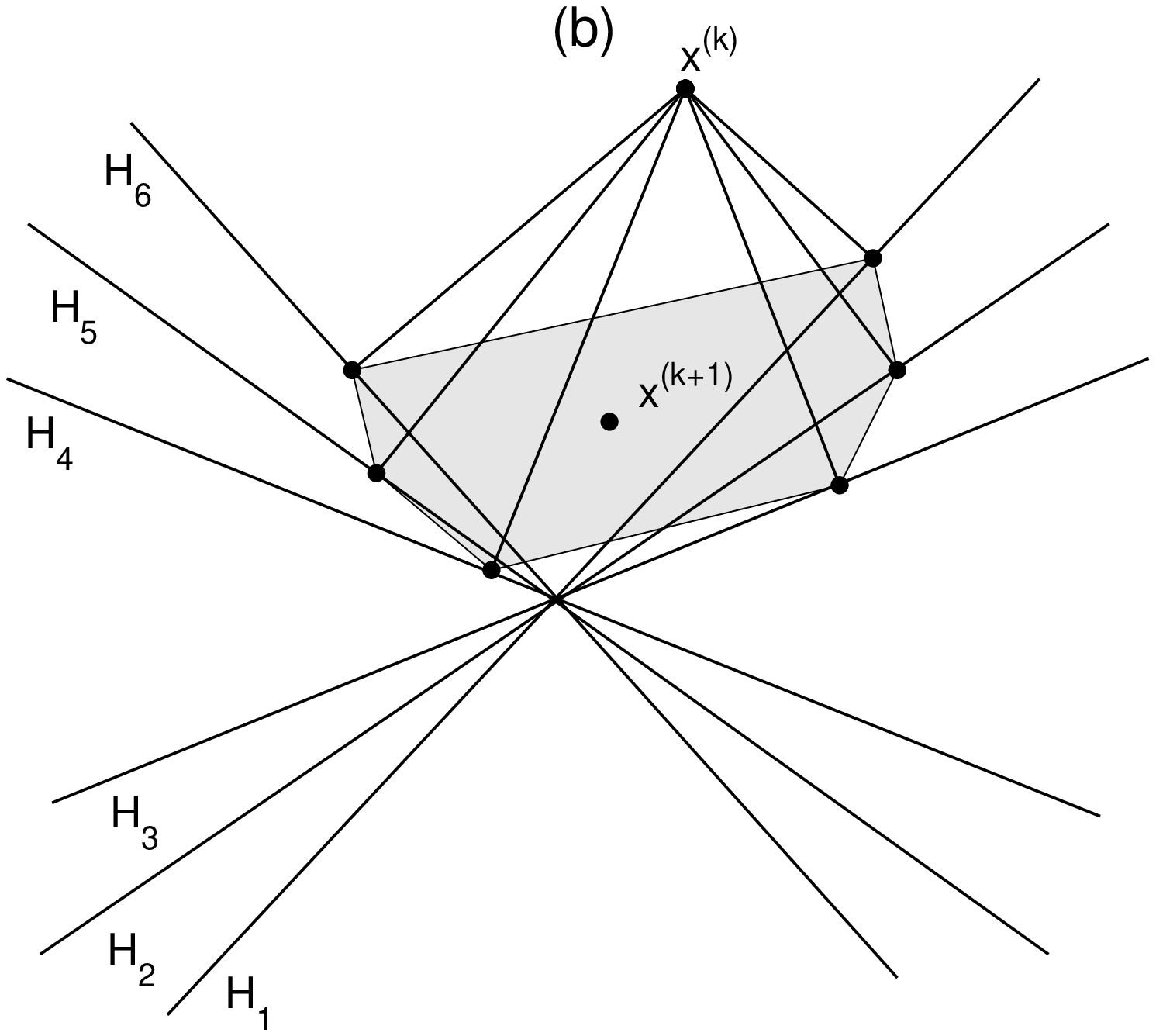}%
}
\\%
{\includegraphics[
height=1.8568in,
width=2.0358in
]%
{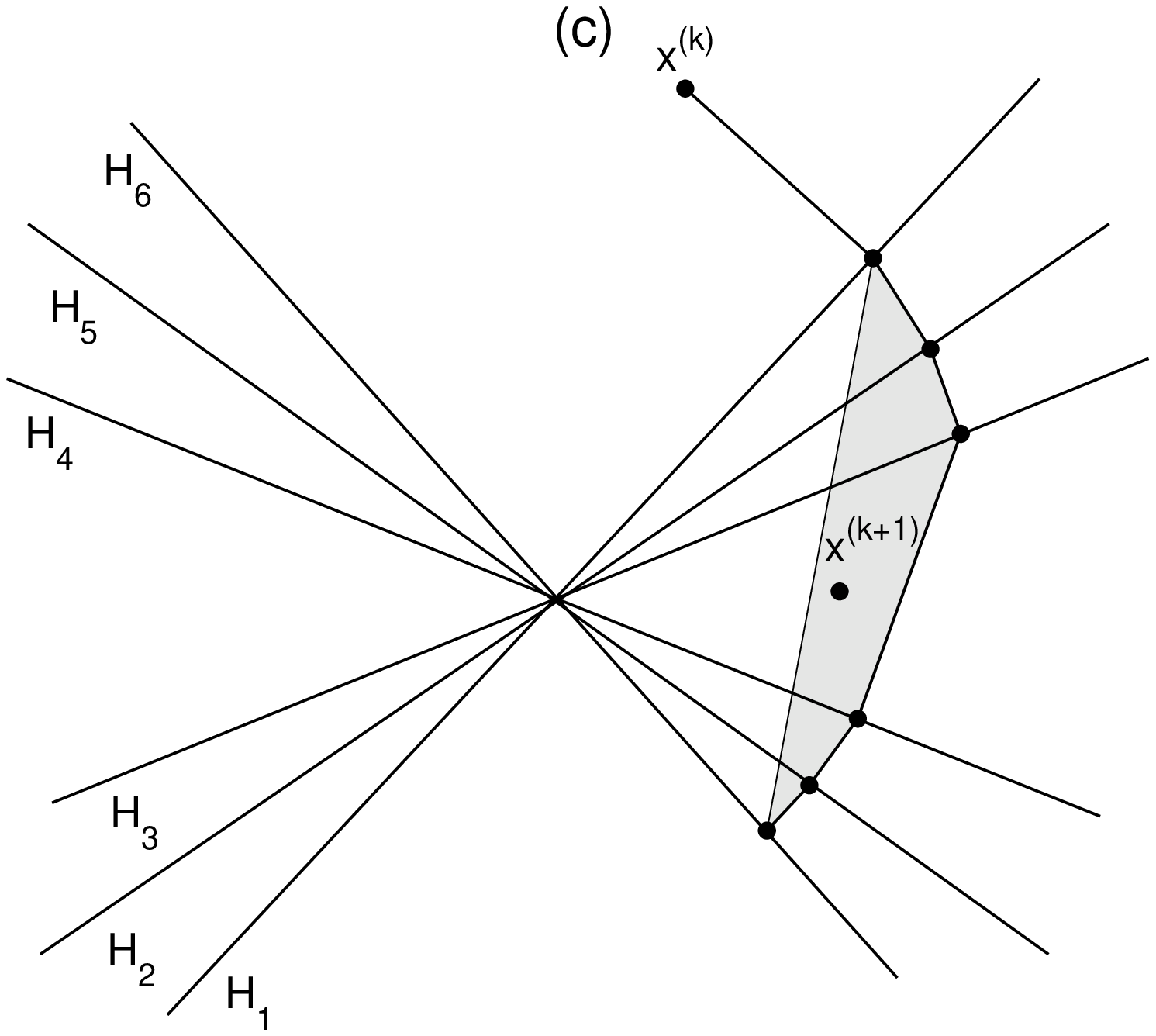}%
}
&
{\includegraphics[
height=1.8568in,
width=2.0358in
]%
{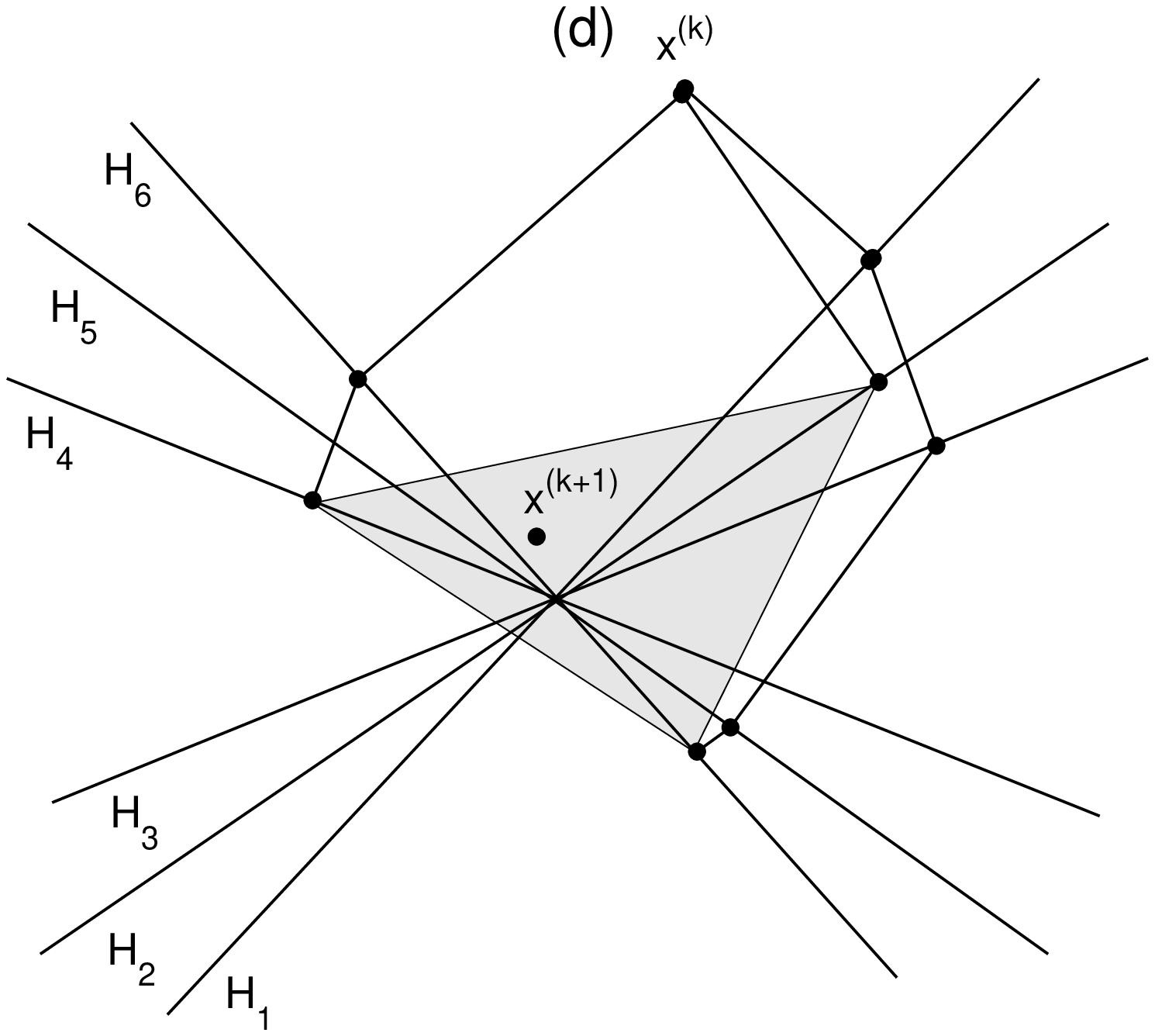}%
}
\end{array}
\]
\caption{(a) Sequential projections. (b) Fully simultaneous projections. (c) Averaging of sequential projections. (d) String-averaging. (Reproduced from \protect\cite{ceh01}).}
\end{figure}

At the extremes of the \textquotedblleft spectrum\textquotedblright\ of possible specific algorithms, derivable from the SAP algorithmic scheme by different choices of strings and weights, are the fully sequential method, which uses one set at a time, and the fully simultaneous algorithm, which employs all sets at each iteration.

\section{Specialized BIP and SAP Reconstruction Algorithms\label{sec:bicav-drop-carp}}

Here we briefly describe the BICAV, DROP, CARP and OS-SART algorithms. More details, further references, and experimental results of applying them to pCT image reconstruction are in \cite{scott-et-al-10,scott10,scott-book} and in the works cited below.

\subsection{BICAV: Block-Iterative Component Averaging\label{subsec:bicav}}

The BICAV (block-iterative component averaging) algorithm of \cite{CGG}, which was applied to a problem of image reconstruction from projections, is a block-iterative companion to the CAV (component averaging) method for solving systems of linear equations \cite{cgg01}. In these methods the sparsity of the matrix is explicitly used when constructing the iteration formula. Using this new scaling considerable improvement was observed compared to traditionally scaled iteration methods. Here is a description of this approach.

In Cimmino's simultaneous projections method (\ref{eq:simultaneous}) with relaxation parameters and with equal weights $w_{i}=1/I$, the next iterate $x^{k+1}$ is the average of the orthogonal projections of $x^{k}$ onto the hyperplanes $H_{i}$ defined by the $i$-th row of the linear system $Ax=b$ and has, for every component $j=1,2,\ldots,J$, the form%
\begin{equation}
x_{j}^{k+1}~=~x_{j}^{k}+\frac{\lambda_{k}}{I}\sum_{i=1}^{I}\frac{b_{i}-\langle
a^{i},x^{k}\rangle}{\left\Vert a^{i}\right\Vert _{2}^{2}}a_{j}^{i}%
\,,\label{2.1.6}%
\end{equation}
where $a^{i}$ is the $i$-th column of the transpose $A^{T}$ of $A$, $b_{i}$ is the $i$-th component of the vector $b,$ and $\lambda_{k}$ are relaxation parameters. When the $I\times J$ system matrix $A=(a_{j}^{i})$ is sparse, only a relatively small number of the elements $\{a_{j}^{1},a_{j}^{2},\ldots ,a_{j}^{I}\}$ of the $j$-th column of $A$ are nonzero, but in (\ref{2.1.6}) the sum of their contributions is divided by the relatively large $I$. This observation led \cite{cgg01} to the replacement of the factor $1/I$ in (\ref{2.1.6}) by a factor that depends on the number of nonzero elements in the set $\{a_{j}^{1},a_{j}^{2},\ldots,a_{j}^{I}\}$. For each $j=1,2,\ldots,J$, denote by $s_{j}$ the number of nonzero elements in column $j$ of the matrix $A,$ and replace (\ref{2.1.6}) by%
\begin{equation}
x_{j}^{k+1}~=~x_{j}^{k}+\frac{\lambda_{k}}{s_{j}}\sum_{i=1}^{I}\frac
{b_{i}-\langle a^{i},x^{k}\rangle}{\Vert a^{i}\Vert_{2}^{2}}a_{j}%
^{i}\,.\label{2.1.62}%
\end{equation}

Certainly, if $A$ is sparse then the $s_{j}$ values will be much smaller than $I,$ and therefore, its use will enable larger additive correction terms in (\ref{2.1.62}). However, in order to prove convergence of such a scheme certain additional changes have to be made to (\ref{2.1.62}), see \cite{CGG,cgg01} for details.

The basic idea of the block-iterative CAV (BICAV) algorithm is to break up the system $Ax=b$ into \textquotedblleft blocks\textquotedblright\ of equations and treat each block according to the CAV methodology, passing cyclically over all the blocks. Use $T$ for the number of blocks and, for $t=1,2,\ldots,T,$ denote, as before, the block of indices $B_{t}\subseteq\{1,2,\ldots,I\}$, by $B_{t}=\{i_{1}^{t},i_{2}^{t},\ldots,i_{n(t)}^{t}\}$, where $n(t)$ is the number of elements in $B_{t}$, such that every element of $\{1,2,\ldots,I\}$ appears in at least one of the sets $B_{t}$. 

For $t=1,2,\ldots,T$, let $A_{t}$ denote the matrix formed by taking all the rows of $A$ whose indices belong to the block of indices $B_{t}$, i.e.,
\begin{equation}
A_{t}:=\left[
\begin{array}
[c]{c}%
a^{\displaystyle i_{1}^{t}}\\
a^{\displaystyle i_{2}^{t}}\\
\vdots\\
a^{\displaystyle i_{n(t)}^{t}}%
\end{array}
\right]  ,~~t=1,2,\ldots,T.\label{At}%
\end{equation}

The iterative step of the BICAV algorithm, developed and experimentally tested in \cite{CGG}, uses, for every block index $t=1,2,\ldots,T,$ orthogonal projections, though each uses a different relaxation parameter, see \cite{CGG} for details. The $\{t(k)\}_{k=0}^{\infty}$ is a control sequence according to which the $t(k)$-th block is chosen by the algorithm to be acted upon at the $k$-th iteration, thus, $1\leq t(k)\leq T$, for all $k\geq0.$ The real numbers $\{\lambda_{k}\}_{k=0}^{\infty}$ are user-chosen relaxation parameters. Finally, let $s_{j}^{t}$ be the number of nonzero elements $a_{j}^{i}\neq0$ in the $j$-th column of $A_{t}$. With these notions at hand the BICAV algorithm is as follows.

\begin{algorithm}
\textbf{Block-iterative component averaging (BICAV)}

\textbf{Initialization: }$x^{0}\in R^{J}$ is arbitrary.

\textbf{Iterative Step: }Given $x^{k},$ compute $x^{k+1}$ by using, for
$j=1,2,\ldots,J,$ the formula:%
\begin{equation}
x_{j}^{k+1}~=~x_{j}^{k}+\lambda_{k}\sum_{i\in B_{t(k)}}\frac{b_{i}-\langle
a^{i},x^{k}\rangle}{\sum_{\ell=1}^{J}s_{\ell}^{t(k)}(a_{\ell}^{i})^{2}}a_{j}%
^{i}\,,\label{eq:bicav}%
\end{equation}
where $\lambda_{k}$ are relaxation parameters, $\{s_{\ell}^{t}\}_{\ell=1}^{J}$
are as defined above, and the control sequence is cyclic, i.e.,
$t(k)=k\operatorname*{mod}T+1,$ for all $k\geq0.$
\end{algorithm}

CAV is the BICAV algorithm described in Eq. (\ref{eq:bicav}) for the special case of a single block containing all equations.

\subsection{DROP: Diagonally Relaxed Orthogonal Projections\label{subsec:drop}}

DROP was developed in order to study and extend Eq. (\ref{2.1.62}) which, as noted above, already appeared in \cite{cgg01}.  In fact, the fully simultaneous DROP was the same as Eq. (\ref{2.1.62}), but with weights $w_i$ added in the summation; see p. 476 of \cite{cehn08}. This was done in \cite{cehn08}, where details and an experimental demonstration are given, yielding the following algorithm. Here blocks and the matrix partition are as in the previous subsection, and the union of all blocks must contain all equations but they may have some common indices.

\begin{algorithm}
\textbf{Diagonally relaxed orthogonal projections (DROP)}

\textbf{Initialization: }$x^{0}\in R^{J}$ is arbitrary.

\textbf{Iterative Step: }Given $x^{k},$ compute $x^{k+1}$ by the formula:%
\begin{equation}
x^{k+1}~=~x^{k}+\lambda_{k}U_{t(k)}\sum_{i\in B_{t(k)}}\frac{b_{i}-\langle
a^{i},x^{k}\rangle}{(a_{\ell}^{i})^{2}}a_{j}^{i}\,,\label{eq:drop}%
\end{equation}
where $\lambda_{k}$ are relaxation parameters, the control sequence is cyclic,
i.e., $t(k)=k\operatorname*{mod}T+1,$ for all $k\geq0,$ and the diagonal
$J\times J$ matrices $U_{t}$ are given, for every $t=1,2,\ldots,T,$ by%
\begin{equation}
U_{t}=\operatorname*{diag} \left(\min(1,1/s_{\ell}^{t} \right),
\end{equation}
with $\{s_{\ell}^{t}\}_{\ell=1}^{J}$ defined as in Subsection
\ref{subsec:bicav} above. 
\end{algorithm}

\subsection{CARP: Component Averaged Row Projections\label{subsec:carp}}

The CARP algorithmic scheme of \cite{carp} is of the general string-averaging variety since it falls under the general definition of \cite[Algorithmic Scheme, page 102]{ceh01}. It resembles the SAP algorithm of Subsection \ref{subsec:sap} with the difference being in the step that combines the strings' end-points $S_{t}(x^{k})$, for all $t=1,2,\dots,T,$ of (\ref{eq:strings}). Contrary to the convex combination used by SAP in (\ref{as2}), the strings' end-points combination is done in CARP as follows.

Given a family of blocks $\{B_{t}\}_{t=1}^{T}$ defined as in Subsection
\ref{subsec:bicav} wherein blocks may overlap, meaning that some equations can belong to 2 or more blocks, define for every $j=1,2,\dots,J,$ the index set%
\begin{equation}
I_{j}=\{t\mid1\leq t\leq T\text{ and }\displaystyle a_{j}^{\displaystyle
i_{q}^{t}}\neq0\text{ for some }1\leq q\leq n(t)\},
\end{equation}
meaning that for the coordinate $j,$ all block indices $t$ of blocks in which
at least one equation has a non-zero coefficient for $x_{j}$ are in the index
set $I_{j}.$ Denoting by $\psi_{j}$ the number of elements in $I_{j},$ the
CARP strings' end-points combination is done by%
\begin{equation}
x_{j}^{k+1}=(1/\psi_{j})\sum_{t\in I_{j}}^{T}(S_{t}(x^{k}))_{j},\text{ \ for
all }j=1,2,\dots,J.
\end{equation}

CARP was originally developed as a domain decomposition method to solve linear systems arising from partial differential equations. Mathematically, CARP is actually ART in some superspace. When each block in CARP (which we call string under the SAP algorithmic paradigm) consists of a single equation, it reduces to Eq. (\ref{2.1.62}), which, as noted above, is also the (unweighted) fully simultaneous DROP. CARP was implemented in \cite{Karonis2013} on a cluster of CPUs and GPUs and achieved \textquotedblleft clinical\textquotedblright reconstruction times on large pCT problems. To the best of our knowledge, this is the only iterative projection method to achieve that to date.

\subsection{OS-SART: Ordered Subsets Simultaneous Algebraic Reconstruction Technique}
Andersen and Kak \cite{aha84} developed a block-iterative technique called the simultaneous algebraic reconstruction technique (SART). They suggested the use of SART with blocks, which the authors called \textquotedblleft subsets,\textquotedblright made up of image projection rays from a single projection angle and in doing so, found that SART was able to deal well with noisy data. The convergence of SART was proven in \cite{MJ01}. The algorithm was developed in such a way that it was equally applicable to subsets, or blocks, of any composition as it was to subsets composed of rays from a single projection angle. This block-iterative form, called ordered subsets simultaneous algebraic reconstruction technique (OS-SART) by Wang and Jiang \cite{gw04}, is as follows. Here again blocks and the matrix partition are as in subsection \ref{subsec:bicav}.

\begin{algorithm}
	\textbf{Ordered subsets simultaneous algebraic reconstruction technique (OS-SART)}
	
	\textbf{Initialization: }$x^{0}\in R^{J}$ is arbitrary.
	
	\textbf{Iterative Step: }Given $x^{k},$ compute $x^{k+1}$ by the formula:%
	\begin{equation}
	x^{k+1}~=~x^{k}+\lambda_{k} \left(\frac{1}{\sum_{i\in B_{t(k)}} a^i_j}\right) \sum_{i\in B_{t(k)}} \frac{b_{i}-\langle a^{i},x^{k}\rangle}{\sum_{\ell = 1}^J a_{\ell}^{i}}a_{j}^{i}\,,\label{eq:ossart}%
	\end{equation}
	where $\lambda_{k}$ are relaxation parameters, the control sequence is cyclic,
	i.e., $t(k)=k\operatorname*{mod}T+1,$ for all $k\geq0,$.
\end{algorithm}

\subsection{Comparison of Algorithms with Proton CT Data}
\begin{figure}[t]
\begin{center}
\includegraphics[width=1.0\textwidth]{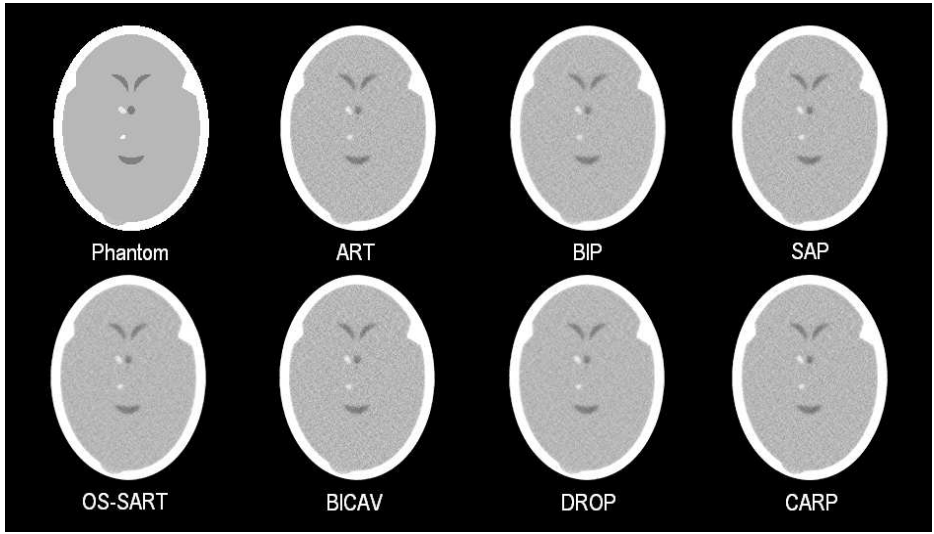}
\caption{Reconstructions of the Herman \cite{H80} head phantom with virtual pCT data. Images presented correspond to a minimum average difference between reconstructed values and known phantom values within 10 iterations. Images were reconstructed with 60 subsets (with the exception of the fully sequential ART) and the optimal relaxation parameter as shown in Fig. \ref{fig:rel_error}. Reproduced from \protect\cite{scott-book} with permission.}
\label{fig:recons}
\end{center}
\end{figure}

\begin{figure}[htp]
\begin{center}
\includegraphics[width=1.0\textwidth]{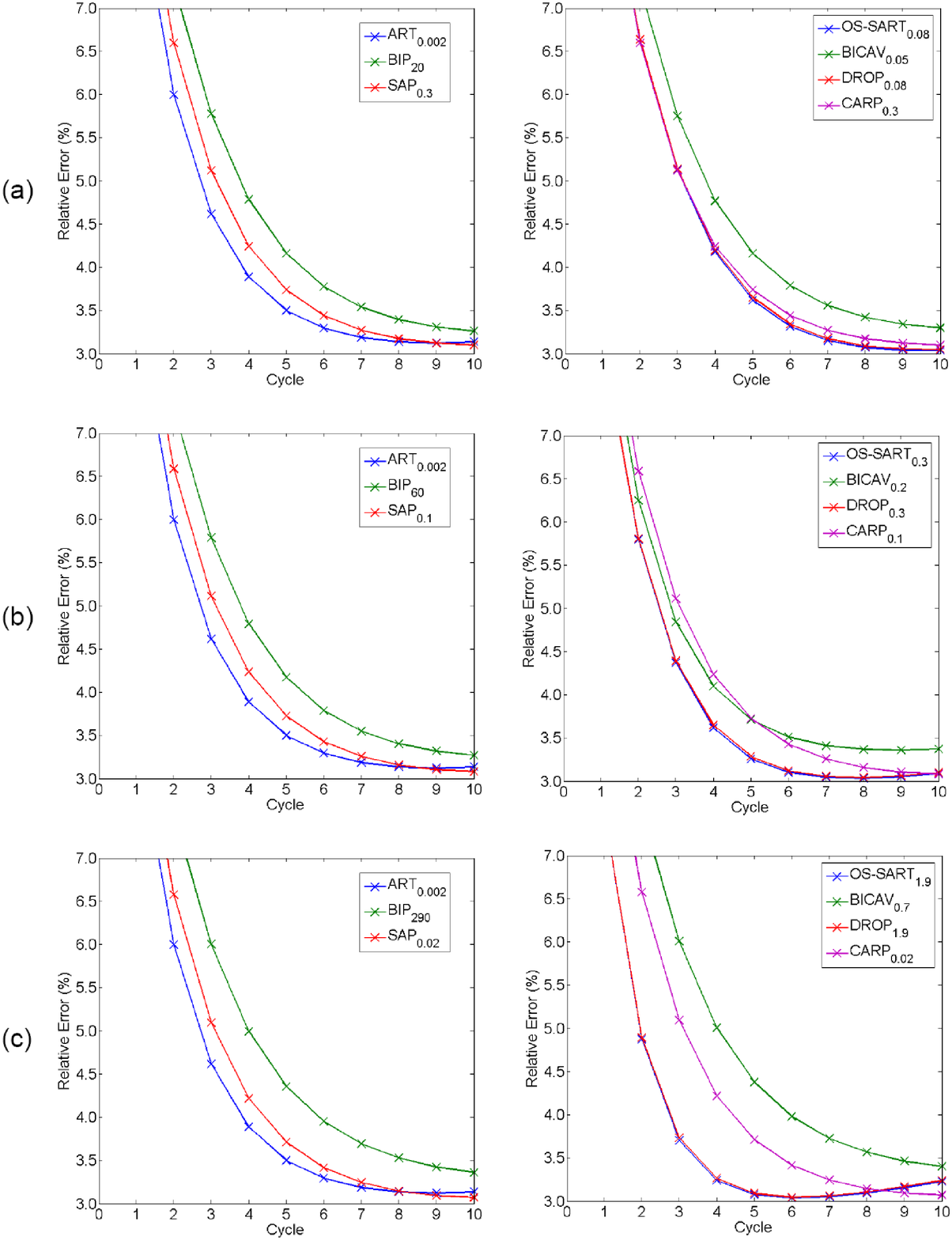}
\caption{Relative error as a function of iteration number for the algorithms presented with (a) 12 (b) 60 and (c) 180 blocks or strings, as appropriate. ART is always reconstructed sequentially (no blocks or strings). The subscript indicates the optimal relaxation parameter determined by trial-and-error. Images with a minimum relative error are presented in Figure \protect\ref{fig:recons}. Reproduced from \protect\cite{scott-book} with permission.}
\label{fig:rel_error}
\end{center}
\end{figure}

A comparison of the algorithms described above has been performed with Monte Carlo simulated pCT data in \cite{scott-et-al-10,scott-book}. It should be noted, however, that the results presented in \cite{scott-book} were produced with a refined system matrix incorporating the concept of effective mean chord lengths as detailed \cite{snp09}. The object used in the simulations was the Herman head phantom \cite{H80} with major and minor axis dimensions of 17.25 cm and 13 cm, respectively. The image reconstruction grid was 256 $\times$ 256 pixels and 20,000 protons per projection angle (180 projection angles evenly spaced over 360 degrees) were used to reconstruct the image. Figure \ref{fig:recons} shows images reconstructed with each algorithm using at most 10 iterations through the pCT dataset. A quantitative assessment of the images was conducted at each iteration by comparing reconstructed RSPs with known RSPs. The quantity used for comparison is termed the relative error $\epsilon$ and is defined by

\begin{equation}
\epsilon = \frac{\sum_j | x^\prime_j-x_j |}{\sum_j x^\prime_j}
\end{equation}

\noindent where $x^\prime_j$ is the known RSP in pixel or voxel $j$. Results for several different block sizes and optimal relaxation parameters are presented in Figure \ref{fig:rel_error}. 

All images produced qualitatively similar reconstructions when combined with an optimal relaxation parameter. However, it should be noted that the primary importance of pCT is to reconstruct quantitatively correct relative stopping powers to be used in proton therapy treatment planning. Small differences in quantitative performance may lead to large differences in treatment planning accuracy. However, whether the differences in relative error between DROP and CARP, for example, relate to a difference in treatment planning accuracy has not yet been demonstrated. Furthermore, the rate of convergence is an important factor in pCT due to the large number of proton paths that must be processed for the reconstruction. Considering these factors, the authors concluded that the block-iterative structure, and DROP in particular, was well-suited for pCT reconstruction with the dataset used in that study. Further studies should compare the reconstruction algorithms on the basis of treatment planning accuracy.

Figure \ref{fig:recons} demonstrates the effect of inconsistent data on pCT reconstructed images. The salt-and-pepper noise limits the contrast resolution. Total-variation superiorization (TVS) has been proposed as a means for reducing this artefact and can be used in conjunction with any of the iterative algorithms described above. A quantitative assessment of the effect of TVS in conjunction with DROP on pCT spatial and contrast resolution was presented in \cite{scott-et-al-10b}. A more recent study also assessed the use of a superiorization framework when dealing with sparse data projections in pCT \cite{Lee2015}. For an updated bibliography on superiorization and perturbation resilience of algorithms consult \cite{ycpubs}.

\section{Most Likely Path in the System Matrix \label{sec:MLP_sys_matrix}}

Better spatial resolution can be achieved in pCT image reconstruction when MCS is accounted for in the system matrix. In iterative reconstructions, the algorithm must determine the relative contribution of each basis function to the line integral. For iterative reconstruction of pCT data, this equates to a calculation of the length of intersection of each particle path with each basis function. The problem then is how to calculate the length of intersection of the nonlinear MLP with each basis function.

To facilitate conceptualization of the problem, only discrete voxelized basis functions are considered. In X-ray CT, the chord lengths of rays with individual pixels can be relatively easily calculated due to the straight line nature of the radiation \cite{rs85}. This is not the case in pCT where MCS deflects the proton path from a straight line. The MLP provides an analytic approximation to the MCS deflected path and can be calculated at discrete points within a scattering object. By \textquotedblleft stepping\textquotedblright $\space$ through the scattering object, discrete MLP points can be joined to form an approximate proton path. The smaller the distance between steps, the more accurate the estimation of pixel intersection length, but the longer the computation time. Although there have been no studies into optimizing MLP step length for computation time and spatial resolution, a step length of one half the pixel side length has been used effectively in pCT iterative reconstructions \cite{snp09,scott-et-al-10b}.

Calculation of individual step-points with the MLP formalism is computationally expensive. To calculate exact chord lengths, a series of decisions and calculations must be made at each step-point, adding to the pCT reconstruction time. A comparison of two methods for determining the elements of the system matrix $A$ for MLP based pCT was presented in \cite{snp09}. In the first method, exact chord lengths are calculated by joining MLP step-points with straight line segments. In the second method, an analytic description of the mean chord length for a given proton path-reconstruction grid orientation is used to assign elements of the system matrix. The latter approach was termed the \textit{effective mean chord length}. The potential advantages of these approaches in quantitative pCT imaging were investigated by reconstructing a Monte Carlo generated pCT data set and using the ART algorithm. It was concluded that the reconstruction time saved when using the effective mean chord length outweighed the minor quantitative improvement in reconstruction values when using exact chord lengths.

With the effective mean chord length a single chord length is assigned to all pixel intersections along a given proton path, speeding up the reconstruction. This approach is based on the assumption that a large number of protons will traverse the image grid with a given orientation and have a uniform spatial distribution. In this case, deterministic proton path-pixel intersections can be treated in a statistical manner. For further details on implementing the effective mean chord length in pCT image reconstruction, the reader is referred to \cite{snp09} and \cite{be11}.

\section{Summary}
Iterative reconstruction algorithms have a long history in medical imaging. Modern advances in computing technology have ensured that some of the original limitations in terms of computation time and memory usage are no longer prohibitive for clinical applications. Statistical iterative reconstruction has been applied in emission tomography and X-ray CT with the goal of modeling photon counting distributions. This resulted in reduced noise in the reconstructed image relative to images reconstructed with analytical algorithms. While a similar approach can be adopted in pCT, no investigations with straggling modelled energy loss have been performed to date, but may yet prove to be a useful tool in reducing straggling induced noise. Rather, image reconstruction in pCT has adopted the projection methods based iterative reconstruction methodology. The combination of these algorithms with an MLP-based system matrix have shown to improve spatial resolution of the reconstructed image relative to images reconstructed with a straight line path approach. However, due to the stochastic nature of the measured quantities in pCT, noise in the reconstructed image limits contrast resolution. Additional techniques have been employed to improve contrast resolution of algebraic iterative reconstruction \cite{scott-et-al-10b}. 

%

\begin{acknowledgements}
The authors would like to thank Reinhard Schulte for his devoted guidance, insightful collaboration and for the encouragement to undertake the writing of this review paper. The authors also thank Dan Gordon for reading and constructively commenting on an earlier version of the paper. We thank three anonymous referees for their reviews which helped us improve the paper. The work of Y.C. was partially supported by the United States-Israel Binational Science Foundation (BSF) Grant number 2013003.
\end{acknowledgements}

\bibliographystyle{ieeetr}
\bibliography{pCT_bib-yair-020815}   


\end{document}